\documentclass[aps,pra,twocolumn,superscriptaddress,groupedaddress]{revtex4}
\usepackage{amssymb} \usepackage{color,graphicx} \usepackage{amsmath}
\usepackage{booktabs}
\usepackage{amsbsy} \usepackage{amsthm} \usepackage{bbm}
\usepackage{bm,bbm} \usepackage{float} \usepackage{braket}
\usepackage{placeins}
\usepackage[colorlinks=true,citecolor=blue,linkcolor=red,urlcolor=red]{hyperref}
\usepackage[sort&compress]{natbib}
\usepackage{comment}
\usepackage{mathtools}
\usepackage{dcolumn,booktabs}
\newcolumntype{d}[1]{D..{#1}}

\usepackage{array,amssymb}
\newcolumntype{P}[1]{>{\centering\arraybackslash}p{#1}}
\newcolumntype{M}[1]{>{\centering\arraybackslash}m{#1}}

\newcommand{\bq}{\begin{equation}} \newcommand{\eq}{\end{equation}}
\newcommand{\bqn}{\begin{equation*}} \newcommand{\eqn}{\end{equation*}}
\newcommand{\bqali}{\begin{equation}\begin{aligned}}
\newcommand{\eqali}{\end{aligned}\end{equation}}

\renewcommand\k{{\bf k}}
\renewcommand\r{{\bf r}}

\newcommand\z{{\bf z}}

\newcommand\x{{\bf x}}
\newcommand\y{{\bf y}}

\graphicspath{{../Figures/}}

\newcommand{\dd}{\text{d}}
\newcommand{\xtilde}{{\raise.17ex\hbox{$\scriptstyle\sim$}}}
\providecommand{\norm}[1]{|#1|}
\providecommand{\ket}[1]{\lvert #1 \rangle}
\providecommand{\ave}[1]{\left\langle  #1 \right\rangle}
\providecommand{\bra}[1]{\langle #1 \lvert}
\begin{document}

\author{J.L. Gaona-Reyes}
\email{joseluis.gaonareyes@phd.units.it}
\affiliation{Department of Physics, University of Trieste, Strada Costiera 11, 34151 Trieste, Italy}
\affiliation{Istituto
Nazionale di Fisica Nucleare, Trieste Section, Via Valerio 2, 34127 Trieste,
Italy}

\author{M. Carlesso}
\affiliation{Department of Physics, University of Trieste, Strada Costiera 11, 34151 Trieste, Italy}
\affiliation{Istituto
Nazionale di Fisica Nucleare, Trieste Section, Via Valerio 2, 34127 Trieste,
Italy}

\author{A. Bassi}
\affiliation{Department of Physics, University of Trieste, Strada Costiera 11, 34151 Trieste, Italy}
\affiliation{Istituto
Nazionale di Fisica Nucleare, Trieste Section, Via Valerio 2, 34127 Trieste,
Italy}

\title{Gravitational interaction through a feedback mechanism}

\date{\today}
\begin{abstract}  
We study the models of Kafri, Taylor and Milburn (KTM) and Tilloy and Di\'osi (TD), both of which implement gravity between quantum systems through a continuous measurement and feedback mechanism. The first model is for two particles, moving in one dimension, where the Newtonian potential is linearized. The second is applicable to any quantum system, within the context of Newtonian gravity. We address the issue of how to generalize the KTM model for an arbitrary finite number of particles. We find that the most straightforward generalisations  are either inconsistent or are ruled out by experimental evidence. We also show that the TD model does not reduce to the KTM model under the approximations which define the latter model. 
We then argue that under the simplest conditions, the TD model is the only viable implementation of a full-Newtonian interaction through a continuous measurement and feedback mechanism.\end{abstract}
\maketitle

\section{Introduction}

Gravity is very well described by General Relativity in terms of space-time deformations caused by mass and energy \cite{Tosto2012}.  All experiments so far have confirmed the theory, up to the  direct detection of gravitational waves by the LIGO collaboration \cite{LIGO2016}, although the open problems in Cosmology with dark energy \cite{Peebles2003} and dark matter \cite{Barack2019} might eventually call for a different description of gravity. 

For decades the scientific community has worked towards formulating a quantum theory of gravity, and many important results have been achieved, yet at present a conclusive answer has not been reached \cite{Kiefer2007, Hossenfelder2011}. 
The lack of a fully consistent quantum theory of gravity has opened  the possibility that gravity might not be fundamentally quantum. The scientific community  considered several times the option that gravity might be ultimately classical, and this originated a dispute whether this assumption is compatible with the quantum nature of matter or not \cite{Feynman1995, Diosi2000, Carlip2001, Huggett2001, Peres2001, Wuthrich2005, Rothman2006, Carlip2008, Dyson2013, Ahmadzadegan2016, Bassi2017, Tilloy2018B, Carney2019, Kumar2020, Donadi2020}, until quite recently, when fully consistent models of Newtonian gravity have been formulated, where matter is quantum and gravity is classical \cite{Diosi1989, Penrose1996, Diosi2011, Bowen2016, Khosla2017, Altamirano2017B, Khosla2018, Altamirano2018, Tilloy2018, Tilloy2019}.
Two of such models were proposed by Kafri, Taylor and Milburn (KTM)~ \cite{Kafri2014}, and by Tilloy and Di\'osi (TD)~\cite{Tilloy2017}.

In this work we study these two models, which implement Newtonian gravity through a continuous measurement and feedback mechanism, whose detailed description is provided below. The first model refers to two particles moving only in one direction, and the Newtonian potential is linearized. The second model applies to any non-relativistic quantum system, and the full Newtonian interaction is considered.  

We address the issue of how to generalize the KTM model for a system of $N$ particles (with $N>2$), taking into account previous results in the literature \cite{Altamirano2018, Kafri2015}. We find that the most straightforward generalisations  are either inconsistent or are ruled out by experimental evidence. 

For the TD model, we analyse the requirements for regularizing the dynamics and we explicitly derive its conditions. In particular, we construct a family of smearing functions for the case of local operations and classical communication (LOCC) dynamics, which is described below. Then, in the appropriate limit, we compare the TD and the KTM models, finding that they predict quantitatively different decoherence effects, and thus concluding that the TD model is not a generalization of the KTM model, although it is built on the same ideas. 

We also address the issue of how to construct a full-Newtonian interaction within a continuous measurement and feedback framework. We argue that the TD model is the only viable one within the simplest form of implementing the feedback mechanism.

\section{Kafri-Taylor-Milburn model} \label{SectKTM}

We review the KTM model; this will allow to set the stage and introduce the key elements for the following discussion. 

The  model \cite{Kafri2014} makes the assumption that Newtonian gravity is fundamentally classical. In order to be consistent with a quantum description of matter, the classical interaction is implemented through a two-step mechanism. The first step is a weak continuous measurement \cite{Jacobs2014} of the positions $\hat x$ of each mass. Then, the (classical) outcome of the position measurement of each mass is coupled to the position of the other mass through a Newtonian potential \cite{Benenti2019,Koashi2016}. This second step corresponds to the implementation of a feedback dynamics. Since the measurement of the positions of the masses has an intrinsic error, the evolution of the system will be characterized by unavoidable noisy dynamics. Thus, this two-step mechanism leads to a decoherence mechanism along-side the desired effective {Newtonian} gravitational attraction between different masses \cite{Bowen2016}.

\begin{figure}[t]
\includegraphics[width=0.8\linewidth]{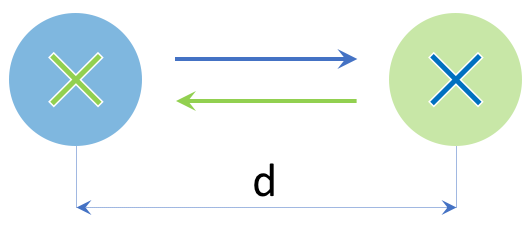}
\caption{Graphical representation of the KTM model. Two particles are initially placed at a  distance $d$ with respect to each other. The position of each particle is measured by the other particle. %
The corresponding measurement record $r_\alpha$ is used to implement a classical {Newtonian} gravitational interaction through a feedback evolution. Here, the measurement is represented with $\bigtimes$ whose color matches the particle performing the measurement, while the measurement record $r_\alpha$ is spread as indicated by the arrows whose color matches the measured particle.
}
\label{KTMfig}
\end{figure}

To be quantitative, KTM considered a system composed of two masses $m_1$ and $m_2$, which are harmonically suspended at an initial distance $d$, as shown in Figure~\ref{KTMfig}, and coupled through gravity, which will be accounted for as presented here below. Since there are only two masses, the problem can be fully studied in one dimension. Assuming that the harmonic trap is sufficiently strong and thus the fluctuations of the masses are small with respect to $d$, one can Taylor expand the {Newtonian} gravitational interaction up to the second order in the relative displacement. Then, with a suitable choice of coordinates, the Hamiltonian of the system reads $\hat{H}=\hat{H}_0+\hat{H}_{\text{grav}}${, where $\hat{H}_0=\sum_{\alpha=1}^2 {\hat{p}_\alpha^2}/{2 m_\alpha}+\frac{1}{2}m_\alpha \Omega_\alpha^2 \hat{x}_\alpha^2$ is the Hamiltonian of a pair of harmonic oscillators}, while $\hat{H}_{\text{grav}}$ describes the linearized interaction due to gravity: 
\begin{equation}
\hat{H}_{\text{grav}}=K\hat{x}_1 \hat{x}_2, \label{HamiltonianKTM}
\end{equation}
where $K={2Gm_1 m_2}/{d^3}$, with $G$ denoting the gravitational constant. The goal of the KTM model is to replace $\hat{H}_{\text{grav}}$, which is quantum in the sense that it depends on the position operators of the two particles,  with the two-step mechanism above described: \textit{i)} measurement of the positions and \textit{ii)} implementation of the feedback dynamics. 

\textit{i) Position measurements. --} The first step is a weak continuous measurement of the positions of the two masses.
According to the standard formalism \cite{Jacobs2014}, the variation of the state {$\ket{\psi}$} due to such a measurement is given by
\bqali
{(\dd \ket{\psi})_{\text{m}}} = &\sum_{\alpha=1}^2\left( - \frac{\gamma_\alpha}{8 \hbar^2}  (\hat{x}_\alpha-\ave{\hat{x}_\alpha})^2 \dd t \right.\\
&\left. +\frac{\sqrt{\gamma_\alpha}}{2 \hbar} (\hat{x}_\alpha - \ave{\hat{x}_\alpha})\dd W_{\alpha,t} \right){\ket{\psi}}, \label{measurement}
\eqali
where {$\ave{\hat{x}_\alpha}=\bra{\psi}\hat{x}_\alpha \ket{\psi}$} and the two noises $W_{\alpha,t}$ are standard independent Wiener processes. The parameters
$\gamma_\alpha$ denote the information rate gained by the measurement.

\textit{ii) Feedback dynamics. --} The feedback dynamics is implemented by replacing $\hat{H}_{\text{grav}}$ with the new feedback Hamiltonian 
\begin{equation}
\hat{H}_{\text{fb}}=\chi_{12} r_1 \hat{x}_2+\chi_{21} r_2 \hat{x}_1, \label{KTMHf}
\end{equation}
with $\chi_{12}$ and $\chi_{21}$ denoting real constants yet to be determined. A key element is the measurement record $r_\alpha$, which encodes the classical information about the position of the $\alpha$-th particle \cite{Jacobs2014}:
\begin{equation}
r_\alpha=\ave{\hat{x}_\alpha} + \frac{\hbar}{\sqrt{\gamma_\alpha}} \frac{\dd W_{\alpha,t}}{\dd t}. \label{KTMsign}
\end{equation}
This is a random variable, centered at the expectation value $\ave{\hat{x}_\alpha}$ and with a variance defined by the information gain rate $\gamma_\alpha$ and the Wiener process $W_{\alpha,t}$, which in turn is defined by the relations
\bqali
\mathbb E[\dd W_{\alpha,t}] &=0,\\
\mathbb{E}\left[\dd W_{\alpha,t}\dd W_{\beta,t} \right] &=\delta_{\alpha \beta}\dd t,
\eqali
see Appendix \ref{AppendixA} for further details.
The Hamiltonian $\hat{H}_{\text{fb}}$ leads to the following feedback evolution for the state:
\begin{equation}
{(\dd \ket{\psi})_{\text{fb}}}=-\sum_{\substack{\alpha,\beta=1 \\ \beta \neq \alpha}}^2 \left[\frac{i}{\hbar}r_{\alpha}+\frac{\chi_{\alpha \beta}\hat{x}_{\beta}}{2 \gamma_\alpha} \right]\chi_{\alpha \beta}\hat{x}_\beta \dd t \,{\ket{\psi}}.\label{feedbackS}
\end{equation}
We report its derivation in Appendix~\ref{AppendixA}.

The full dynamics of the state $\ket\psi$ is  then given by the combining the contributions in Eq.~\eqref{measurement} and Eq.~\eqref{feedbackS}. This reads
\begin{equation}
\begin{split}
\dd \ket{\psi}\!&=\!\left\{{-\sum_{\substack{\alpha,\beta=1 \\ \beta \neq \alpha}}^2 \left[\frac{i}{\hbar}r_{\alpha}+\frac{\chi_{\alpha \beta}\hat{x}_{\beta}}{2 \gamma_\alpha} \right]\chi_{\alpha \beta}\hat{x}_\beta}\dd t \right.\\
&\left. +\sum_{\alpha=1}^2\left[-\frac{\gamma_\alpha}{8 \hbar^2}(\hat{x}_\alpha- \ave{\hat{x}_\alpha})^2 \dd t + \frac{\sqrt{\gamma_\alpha}}{2 \hbar}(\hat{x}_\alpha-\ave{\hat{x}_\alpha})  \dd W_{\alpha,t}\right] \right.\\
&\left.-\frac{i}{2 \hbar}\sum_{\substack{\alpha,\beta=1\\ \beta \neq \alpha}}^2 \chi_{\alpha \beta}\hat{x}_\beta(\hat{x}_\alpha-\ave{\hat{x}_\alpha}) \dd t \right\}\ket{\psi}, 
\end{split} \label{KTMfull}
\end{equation} 
where the first line corresponds to the feedback contribution, the second line to that of the continuous measurement, while the third line is the It\^o term arising from their combined effect.
The corresponding master equation reads 
\bqali
\frac{\dd \hat{\rho}}{\dd t}=&-\frac{i}{\hbar}\left[\hat{H}_0,\hat{\rho}\right]-\frac{i}{2 \hbar}\sum_{\substack{\alpha,\beta=1 \\ \beta \neq \alpha}}^2 \chi_{\alpha \beta}\left[\hat{x}_\beta,\left\{\hat{x}_\alpha,\hat{\rho} \right\}\right]\\
&{-\sum_{\alpha=1}^2 \left(\frac{\gamma_\alpha}{8 \hbar^2}+\sum_{\substack{\beta=1 \\ \beta \neq \alpha}}^2 \frac{\chi_{\beta \alpha}^2}{2 \gamma_\beta} \right)\left[\hat{x}_\alpha,\left[\hat{x}_\alpha,\hat{\rho}\right]\right]}, \label{KTMdyn}
\eqali
with $\hat \rho=\mathbb E[\ket\psi\bra\psi]$, where  $\mathbb{E}[ \cdot ]$ denotes the stochastic average; in Eq.~\eqref{KTMdyn} we also included the free evolution described by $\hat H_0$. 

To correctly mimic the  gravitational interaction one sets $\chi_{12}=\chi_{21}=K$, and the master equation becomes
\bqali
\frac{\dd \hat{\rho}}{\dd t}=&-\frac{i}{\hbar}\left[\hat{H}_0+K \hat{x}_1 \hat{x}_2,\hat{\rho}\right]\\
&-\sum_{\alpha=1}^2 \left(\frac{\gamma_\alpha}{8 \hbar^2}+\sum_{\substack{\beta=1\\ \beta \neq \alpha}}^2\frac{K^2}{2 \gamma_\beta} \right)\left[\hat{x}_\alpha,\left[\hat{x}_\alpha,\hat{\rho}\right]\right].\label{KTMrecov}
\eqali 
Hence, one recovers, at the level of the master equation, the quantum gravitational interaction $\hat{H}_{\text{grav}}$ in Eq.~\eqref{HamiltonianKTM}. In this way, the KTM prescription retrieves the standard Newtonian quantum  gravitational interaction through a classical communication channel. 
The price to pay is to have decoherence effects, whose strength is determined by the parameters $\gamma_\alpha$. In the particular case of two equal masses $m_1=m_2$, it is reasonable to assume that the measurement processes have the same rate \cite{Kafri2014}, thus we set $\gamma=\gamma_1=\gamma_2$. This is a free parameter of the model, which can be fixed by looking for a minimum.  
The particular structure of the decoherence terms in Eq.~\eqref{KTMrecov} allows to perform such a minimization, after which the second line of Eq.~\eqref{KTMrecov} becomes
\bq
-\frac K{2\hbar}{\sum_{\alpha=1}^2} [\hat x_\alpha,[\hat x_\alpha,\hat \rho]],
\label{eq.decoherenceDP}
\eq
and corresponds to an information gain rate equal to $\gamma_\text{min}=2 \hbar K$ \cite{Kafri2014}.

In summary, the KTM model implements a local operation and classical communication (LOCC) dynamics \cite{Chitambar2014, Marshman2020}, where the local operation is provided by the continuous measurement of the positions, while the feedback dynamics works as a classical communication \cite{Horodecki2009}. Such a LOCC dynamics simulates the action of a Newtonian quantum  gravitational field in the sense specified above, paying the price of having a decoherence mechanism affecting the system dynamics.

\section{Linearized-gravity generalization of the KTM model}\label{sec.3}

The KTM model describes the Newtonian gravitational interaction of two particles only. A natural question is how to generalize it to include an arbitrary finite number of particles: this is the subject of this section. We will keep gravity at linear order. 

Two  generalizations naturally follow from the original proposal: the first ones assumes that the position of each mass is measured by each of the  other masses (pairwise measurement); the second one assumes a single, universal measurement of the position of each mass. Then the measurement records are used to implement the feedback dynamics consistently.

\subsection{Pairwise approach}

The pairwise approach was first proposed  by Altamirano  \textit{et al.}~\cite{Khosla2017,Altamirano2017,Altamirano2018},    where they considered two bodies of $N_1$ and $N_2$ constituents, moving in one dimension. We review the model, and at the same time we generalize it to a arbitrary configuration of particles in three dimensions. 

The Taylor expansion of the  many body Newtonian potential reads
\begin{equation}
\hat{V}  \approx \sum_{\alpha=1}^N\hat{Y}_\alpha +\frac{1}{2}\sum_{\substack{\alpha,\beta=1 \\ \beta \neq \alpha}}^N \sum_{l,j=1}^3 K_{\alpha \beta lj}\hat{x}_{\alpha l}\hat{x}_{\beta j}, \label{Altpot}
\end{equation}
where the Greek indices  $\alpha,\beta$ denote the particles and the Latin indices $l,j$ denote the directions in space. The single particle operator $\hat{Y}_\alpha$ is a second-order polynomial  of the position operator $\hat{\x}_\alpha$, which is not relevant because it can be re-absorbed with a proper redefinition of the variables,  while the second term gives the Newtonian potential at linear order, with the coefficients $K_{\alpha \beta l j}$ defined as follows:
\begin{equation}\label{defK3d}
K_{\alpha \beta lj}=Gm_\alpha m_\beta \left[\frac{3 d_{\alpha \beta l}d_{\alpha \beta j}}{|\mathbf{d}_{\alpha \beta}|^5}-\frac{\delta_{lj}}{|\mathbf{d}_{\alpha \beta}|^3} \right],
\end{equation}
where the vector $\mathbf{d}_{\alpha \beta}$ joins the positions of the two masses. This is the generalization of $K$ introduced in Eq.~\eqref{HamiltonianKTM}.

We now apply the two-step mechanism outlined before. The variation of the wavefunction due to the continuous measurements of the positions $\hat{x}_{\alpha l}$ is described by:
\bqali
{(\dd \ket{\psi})_{\text{m}}}&=\sum_{\substack{\alpha,\beta=1 \\ \beta \neq \alpha}}^N \sum_{l,j=1}^3 \left(-\frac{\gamma_{\alpha \beta lj}}{8 \hbar^2}\left(\hat{x}_{\alpha l}-\ave{\hat{x}_{\alpha l}}\right)^2 \dd t \right.\\
&\left.+\frac{{\sqrt{\gamma_{\alpha \beta lj}}}
}{2 \hbar}\left(\hat{x}_{\alpha l}-\ave{\hat{x}_{\alpha l}} \right) \dd W_{\alpha \beta lj,t}\right){\ket{\psi}},
\label{eq:dfgh}
\eqali
where the parameters $\gamma_{\alpha \beta lj}$ are the information gain rates of the measurements, and the  noises $ W_{\alpha \beta lj,t}$ are standard independent Wiener processes, satisfying
\bqali
\mathbb{E}\left[\dd W_{\alpha \beta lj,t}\right]&=0\\
\mathbb{E}\left[\dd W_{\alpha \beta lj,t}\dd W_{\alpha' \beta' l' j',t}\right]&=\delta_{\alpha \alpha'}\delta_{\beta \beta'}\delta_{ll'}\delta_{jj'}\dd t.
\eqali
The corresponding measurement records read
\begin{equation}\label{meas.rec.KTM.gen}
r_{\alpha \beta lj}=\ave{\hat{x}_{\alpha l}}+\frac{\hbar}{\sqrt{\gamma_{\alpha \beta lj}}}\frac{\dd W_{\alpha \beta lj,t}}{\dd t}.
\end{equation}
We speak of a pairwise approach because, as we can see from Eq.~\eqref{eq:dfgh}, the position of each particle is measured $9(N-1)$ times in the three directions in space, for a total of  $9N(N-1)$ measurement records $r_{\alpha \beta lj}$.
These  embed the information about the position of  particle $\alpha$ along the $l$-th direction, which will be used to generate the gravitational attraction on particle $\beta$ along the $j$-th direction. Different particles will use
\begin{figure}
\includegraphics[width=0.7\linewidth]{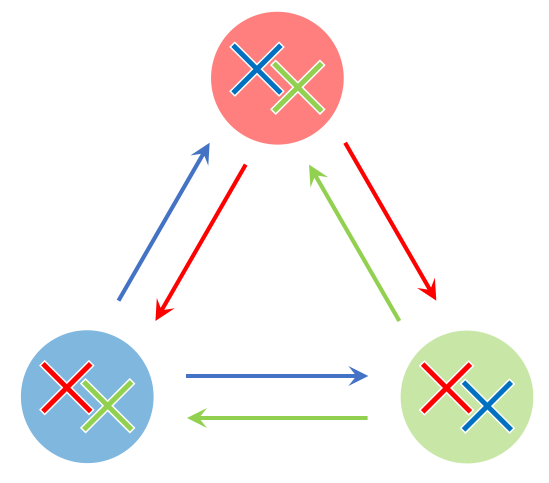}
\caption{Graphical representation of the pairwise KTM generalization for $N=3$ particles. Each particle position is measured by all the other $N-1=2$ particles. Here, the measurement is represented with $\bigtimes$ whose color matches the particle performing the measurement, while the measurement record $\r_{\alpha\beta}$ is broadcasted as indicated by the arrows whose color matches the measured particle.}
\label{fig.Altamirano}
\end{figure}
{\it different} measurement records coming from the same particle to implement the feedback dynamics. See Fig.~\ref{fig.Altamirano} for a graphical representation of the situation.

The feedback Hamiltonian is defined as follows:
\begin{equation}
\hat{H}_{\text{fb}}=\sum_{\substack{\alpha,\beta=1 \\ \beta \neq \alpha}}^N \sum_{l,j=1}^3K_{\alpha \beta lj}r_{\alpha \beta lj}\hat{x}_{\beta j}. \label{Altfeed}
\end{equation}
The corresponding feedback evolution for the wavefunction is given by
\bqali
{(\dd \ket{\psi})_{\text{fb}}}\!=\!- \!\!\! \sum_{\substack{\alpha,\beta=1 \\ \beta \neq \alpha}}^N \sum_{l,j=1}^3 \!\!\left[\frac{i}{\hbar}r_{\alpha \beta lj}\!+\!\frac{K_{\alpha \beta lj}\hat x_{\beta j}}{2 \gamma_{\alpha \beta lj}} \right]\!\!K_{\alpha \beta lj}\hat{x}_{\beta j} \dd t \, {\ket{\psi}}.
\eqali
Following the procedure outlined in the previous section and reported in Appendix~\ref{AppendixA}, we arrive at the following master equation for the combined measurement and feedback dynamics:
\bqali
\frac{\dd \hat{\rho}}{\dd t}=&-\frac{i}{\hbar}\left[\hat{H}_0,\hat{\rho} \right]-\frac{i}{2 \hbar}\sum_{\substack{\alpha,\beta=1 \\ \beta \neq \alpha}}^N \sum_{l,j=1}^3 K_{\alpha \beta lj}\left[\hat{x}_{\alpha l}\hat{x}_{\beta j},\hat{\rho}\right]\\
&-\sum_{\substack{\alpha,\beta=1 \\\ \beta \neq \alpha}}^N \sum_{l,j=1}^3 \left(\frac{\gamma_{\alpha \beta lj}}{8 \hbar^2}+\frac{1}{2}\frac{K_{\alpha \beta lj}^2}{\gamma_{\alpha \beta lj}} \right)\left[\hat{x}_{\alpha l},\left[\hat{x}_{\alpha l},\hat{\rho}\right]\right] ,
\label{Altamirano}
\eqali
where we absorbed the operators $\hat{Y}_\alpha$ in the Hamiltonian $\hat{H}_0$ {and we assumed that $\gamma_{\alpha \beta lj}=\gamma_{\beta \alpha lj}=\gamma_{\alpha \beta jl}$}. The master equation \eqref{Altamirano}  is a three-dimensional pairwise generalization of the KTM model. 
As before, the unitary evolution (apart from $\hat{H}_0$) describes the  gravitational interaction among the particles. The other terms in the second line   lead to decoherence, 
and can be suitably minimized by fixing an appropriate value of the information rates $\gamma_{\alpha \beta lj}$. Thus, as for the KTM model, one obtains a minimum decoherence coefficient that can be tested experimentally. {Note that Eq.~\eqref{Altamirano} reduces to the KTM master equation \eqref{KTMrecov} for $N=2$ particles. The same holds when one considers
 two subsystems made of respectively $N_1$ and $N_2$ particles, with $N_1+N_2=N$. Indeed, once expressing each of the position operators $\hat \x_{\alpha}$ as the sum of the center-of-mass operator $\hat {\bf X}_\alpha^\sigma$, with $\sigma=1$ or 2, and relative displacement $\delta\hat\x_\alpha^\sigma$, we have that the double commutator in Eq.~\eqref{Altamirano} can be expressed as
\bqali
\left[\hat{x}_{\alpha l},\left[\hat{x}_{\alpha l},\hat{\rho}\right]\right]=\left[\hat X_{\alpha l}^\sigma,\left[\hat X_{\alpha l}^\sigma,\hat{\rho}\right]\right]+\left[\hat X_{\alpha l}^\sigma,\left[\delta\hat x_{\alpha l}^\sigma,\hat{\rho}\right]\right]\\+\left[\delta\hat x_{\alpha l}^\sigma,\left[\hat X_{\alpha l}^\sigma,\hat{\rho}\right]\right]+\left[\delta\hat x_{\alpha l}^\sigma,\left[\delta\hat x_{\alpha l}^\sigma,\hat{\rho}\right]\right].
\eqali
Then, by tracing over the relative degrees of freedom,
\bq
\operatorname{Tr}^\text{\tiny rel}\left(\left[\hat{x}_{\alpha l},\left[\hat{x}_{\alpha l},\hat{\rho}\right]\right]\right)=\left[\hat X_{\alpha l}^\sigma,\left[\hat X_{\alpha l}^\sigma,\hat{\rho}_\text{\tiny CM}\right]\right],
\eq
and one recovers the dynamics in Eq.~\eqref{KTMrecov} for the centers-of-mass of the two subsystems.}\\
This model is mathematically consistent, however it is  experimentally ruled out as discussed in Ref.~\cite{Altamirano2018}.
{Indeed, each of the particles is measured as many times as the number of the other particles present in the system. If one takes the example of the system made of an atom and the entire Earth \cite{Altamirano2018}, then every particle constituting the latter  provides a contribution to the decoherence term in Eq.~\eqref{Altamirano}. For the atom, after tracing over the Earth's degrees of freedom, one obtains the decoherence term in the vertical direction $z$   of motion
\bq\label{earthatom}
-\frac{\mathcal CG m_\text{\tiny atom}M_\oplus}{\hbar R_\oplus^3}[\hat z,[\hat z,\hat \rho_\text{\tiny atom}]],
\eq
after the minimization procedure is applied. Here, $m_\text{\tiny atom}\sim1.4\times 10^{-25}\,$kg is the mass of the $^{87}$Rb atom used in the considered  experiment \cite{Kovachy:2015aa}, $M_\oplus\sim6\times10^{24}\,$kg and $R_\oplus\sim6 \times10^6\,$m are the mass and the radius of Earth, while $\mathcal C\sim0.47$ is a suitable factor accounting for the Earth's geometry \cite{Altamirano2018}. By 
following the analysis in \cite{Altamirano2018}, one finds that Eq.~\eqref{earthatom} predicts a visibility which is more than 80 orders of magnitude smaller than that experimentally measured.}

\subsection{Universal approach}

{Since a pairwise procedure, where every particle measures the others, is experimentally ruled out due to the excessive number of measurements, one needs to consider an alternative \cite{Khosla2017}. Here we consider an {\it universal} approach, where 
the position of each particle is measured only once, and such an information is broadcasted to all the other particles through the 
feedback Hamiltonian. }

\begin{figure}[t!]
\includegraphics[width=0.7\linewidth]{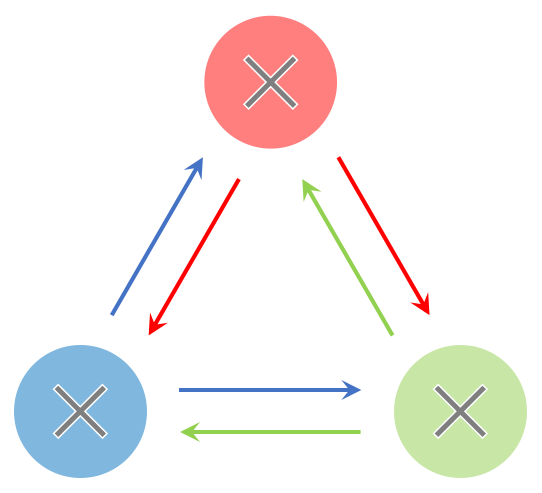}
\caption{Graphical representation of the universal KTM generalization for $N=3$ particles. Each particle position is measured only once, this is represented by the grey $\bigtimes$. The corresponding measurement records $\r_\alpha$ are broadcasted as indicated by the arrows whose color matches the measured particle.}
\label{AltamiranoAlt}
\end{figure}

{Now, we have one measurement record for each of the $N$ particles in each of the three dimensions, of the form 
\bq
r_{\alpha l}=\braket{\hat x_{\alpha l}}+\frac{\hbar}{\sqrt{\gamma_{\alpha l}}}\frac{\dd W_{\alpha l,t}}{\dd t},
\eq
where the noise is characterized by
\bqali
\mathbb E[\dd W_{\alpha l,t}] &=0,\\
\mathbb{E}\left[\dd W_{\alpha l,t}\dd W_{\beta j,t} \right] &=\delta_{\alpha \beta}\delta_{lj}\dd t.
\eqali}
In such a way, once the position of one particle is measured, the other particles receive the same measurement record. A graphical scheme of such a protocol is shown in Fig.~\ref{AltamiranoAlt}. After the continuous measurements, which is described by 
{
\bqali
{(\dd \ket{\psi})_{\text{m}}} = &\sum_{\alpha=1}^N\sum_{l=1}^3\left( - \frac{\gamma_{\alpha l}}{8 \hbar^2}  (\hat{x}_{\alpha l}-\ave{\hat{x}_{\alpha l}})^2 \dd t \right.\\
&\left. +\frac{\sqrt{\gamma_{\alpha l}}}{2 \hbar} (\hat{x}_{\alpha l} - \ave{\hat{x}_{\alpha l}})\dd W_{{\alpha l},t} \right){\ket{\psi}}, \label{measurement}
\eqali}
one implements the  gravitational interaction through the following feedback Hamiltonian
\begin{equation}
\hat{H}_{\text{fb}}=\sum_{\substack{\alpha,\beta=1 \\ \beta \neq \alpha}}^N\sum_{l,j=1}^3K_{\alpha \beta l j}r_{\alpha l} \hat{x}_{\beta j},
\end{equation}
{where $K_{\alpha\beta l j}$ is defined in Eq.~\eqref{defK3d}. The corresponding contribution to the evolution of the wavefunction reads
\bqali
{(\dd \ket{\psi})_{\text{fb}}}\!=\!- \!\! \sum_{\alpha=1}^N \sum_{l=1}^3 \left[\frac{i}{\hbar}r_{\alpha  l}+\sum_{\substack{\epsilon=1 \\ \epsilon \neq \alpha}}^N\sum_{i=1}^3\frac{K_{\alpha \epsilon li}\hat x_{\epsilon i}}{2 \gamma_{\alpha  l}} \right]\\
\times\sum_{\substack{\beta=1 \\ \beta \neq \alpha}}^N\sum_{j=1}^3K_{\alpha \beta lj}\hat{x}_{\beta j} \dd t \, {\ket{\psi}}.
\eqali}
Following the procedure described in Appendix~\ref{AppendixA}, we obtain the following master equation
\bqali
\frac{\dd \hat{\rho}}{\dd t}=&-\frac{i}{\hbar}\left[\hat{H}_0,\hat{\rho}\right]-\frac{i}{2 \hbar}\sum_{\substack{\alpha, \beta=1 \\ \beta \neq \alpha}}^N \sum_{l,j=1}^3K_{\alpha \beta l j}\left[\hat{x}_{\beta j},\left\{\hat{x}_{\alpha l},\hat{\rho}\right\}\right]\\
&-\sum_{\alpha=1}^N\sum_{l=1}^3 \frac{\gamma_{\alpha l}}{8 \hbar^2}\left[\hat{x}_{\alpha l},\left[\hat{x}_{\alpha l},\hat{\rho}\right]\right]\\
&-\sum_{\substack{\alpha,\beta, \epsilon=1 \\ \beta, \epsilon \neq \alpha}}^N \sum_{l,j,i=1}^3\frac{K_{\alpha \beta l j}K_{\alpha \epsilon l i}}{2 \gamma_{\alpha l}}\left[\hat{x}_{\beta j},\left[\hat{x}_{\epsilon i},\hat{\rho}\right]\right]. \label{KTMfake}
\eqali
Differently from Eq.~\eqref{KTMrecov} and Eq.~\eqref{Altamirano}, the decoherence term in Eq.~\eqref{KTMfake} involves the position operators of different particles. This poses a serious problem. To see this, let us consider again the case of two subsystems made of $N_1$ and $N_2$ particles respectively. 
{Now, by splitting $\hat \x_{\beta}$ as the sum of the center-of-mass operator $\hat {\bf X}_\beta^\sigma$, with $\sigma=1$ or 2, and relative displacement $\delta\hat\x_\beta^\sigma$, we have that the last double commutator in Eq.~\eqref{KTMfake} becomes 
\bqali
\left[\hat{x}_{\beta j},\left[\hat{x}_{\epsilon i},\hat{\rho}\right]\right]=\left[\hat X_{\beta j}^\sigma,\left[\hat X_{\epsilon i}^\mu,\hat{\rho}\right]\right]+\left[\hat X_{\beta j}^\sigma,\left[\delta\hat x_{\epsilon i}^\mu,\hat{\rho}\right]\right]\\+\left[\delta\hat x_{\beta j}^\sigma,\left[\hat X_{\epsilon i}^\mu,\hat{\rho}\right]\right]+\left[\delta\hat x_{\beta j}^\sigma,\left[\delta\hat x_{\epsilon i}^\mu,\hat{\rho}\right]\right]
\eqali
Then, by tracing over the relative degrees of freedom, one finds 
\bq\label{eq.strage.double}
\operatorname{Tr}^\text{\tiny rel}\left(\left[\hat{x}_{\beta j},\left[\hat{x}_{\epsilon i},\hat{\rho}\right]\right]\right)=\left[\hat X_{\beta j}^\sigma,\left[\hat X_{\epsilon i}^\mu,\hat{\rho}_\text{\tiny CM}\right]\right],
\eq
where $\mu$ and $\sigma$ do not necessarily coincide. Thus, one has also terms of the form $\left[\hat X_{\beta j}^1,\left[\hat X_{\epsilon i}^2,\hat{\rho}_\text{\tiny CM}\right]\right]$, which do not appear in the KTM master equation \eqref{KTMrecov}. {In Appendix \ref{coefUni} we present an explicit example proving that the corresponding coefficient is non vanishing.} On the contrary, for the case of $N=2$, one recovers exactly the KTM model without additional terms. Indeed, for $N=2$, the constraint $\beta,\epsilon\neq\alpha$ in the last term of Eq.~\eqref{KTMfake} is satisfied only for $\beta=\epsilon$ and $\beta\neq\alpha$. Thus, one does not have double commutators involving position operators of different particles. Therefore, two composite systems do not behave like two point-like particles, whose internal dynamics can be ignored. This inconsistency discards the universal generalization of the KTM model.
}

In the next Section we will consider the model proposed by Tilloy and Diosi~\cite{Tilloy2016,Tilloy2017}, which  consistently describes gravity as a measurement plus feedback interaction for an arbitrary number of particles, also implementing the full Newtonian potential, not only its linear approximation. We will show that it does not reduce to the KTM model in the limit of linearized gravity for two particles.  

\section{Tilloy-Di\'osi model} \label{TDmodsect}

The TD model \cite{Tilloy2016,Tilloy2017} implements a full Newtonian interaction by monitoring the 
mass density of the system. This choice allows a straightforward extension to the case of identical particles, where one expresses the mass density operator as a mass-weighted sum over different species of particles. In this setting, Eq.~\eqref{HamiltonianKTM} is replaced by
\begin{equation}
\hat{H}_{\text{grav}}=\frac{1}{2}\int \dd \mathbf{x} \dd \mathbf{y} \mathcal{V}(\mathbf{x}-\mathbf{y})\hat{\mu}(\mathbf{x})\hat{\mu}(\mathbf{y}), \label{Hgrav}
\end{equation} 
where $\mathcal{V}(\mathbf{x}-\mathbf{y})=-G /\norm{\mathbf{x}-\mathbf{y}}$ is the {full Newtonian} gravitational potential and $\hat{\mu}(\mathbf{x})$ is the mass density operator of the system.  This is the interaction one aims at recovering with the measurement and feedback process.

In analogy with the KTM model, now the mass density $\hat{\mu}(\mathbf{x})$ is continuously measured.  The variation of the wavefunction due to the continuous measurement  is given by
\bqali
&{(\dd \ket{\psi})_\text{m}}=\left[-\frac{1}{8 \hbar^2}\int \dd \mathbf{x} \dd \mathbf{y} \gamma(\mathbf{x},\mathbf{y})\left(\hat{\mu}(\mathbf{x})-\ave{\hat{\mu}(\mathbf{x})}\right)\right.\\
&\times \left(\hat{\mu}(\mathbf{y})-\ave{\hat{\mu}(\mathbf{y})}\right)\dd t  \\
&\left.+\frac{1}{2 {\hbar}}\int \dd \mathbf{x} \left(\hat{\mu}(\mathbf{x})-\ave{\hat{\mu}(\mathbf{x})} \right)\delta \mu_t(\mathbf{x})\dd t \right]{\ket{\psi}}
. \label{TDmeas}
\eqali
This is the analogue of Eq.~\eqref{measurement} in the KTM model. Here, we introduced $\ave{\hat{\mu}(\mathbf{x})}={\bra{\psi}\hat{\mu}(\mathbf{x})\ket{\psi}}$ and the noise $\delta \mu_t(\mathbf{x})$ (the generalization of $\frac{\dd W_t}{\dd t}$ of the KTM model) is characterized by
\bqali
\mathbb{E}[{\delta\mu_t(\mathbf{x})}]&=0,\\
\mathbb{E} \left[\delta \mu_t (\mathbf{x}) \delta \mu_{t'}(\mathbf{y})\right]&=\gamma(\mathbf{x},\mathbf{y})\delta(t-t'), \label{correl}
\eqali
where $\gamma(\mathbf{x},\mathbf{y})$ is a  spatial correlator. The latter is assumed to be symmetric, {and} satisfying $\gamma(\x,\y)=\gamma(\y,\x){=\gamma(\x-\y)}$.

\begin{figure}[t!]
\includegraphics[width=0.7\linewidth]{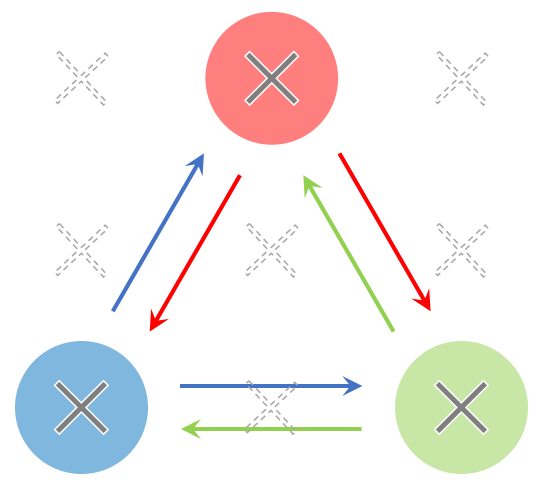}
\caption{Graphical representation of the TD model scheme for $N=3$ particles. The mass-density in each point of space is measured: whether in that particular position there is a particle (grey $\bigtimes$) or not (dashed $\bigtimes$). If a non-zero value of the mass density is found, then the corresponding measurement record $\mu(\x)$ is broadcasted as indicated by the arrow whose color matches the measured particle.}
\label{TDgraph}
\end{figure}

In analogy with Eq.~\eqref{KTMHf}, we introduce the feedback Hamiltonian
\begin{equation}
\hat{H}_\text{fb}=\int \dd \mathbf{x} \dd \mathbf{y} \mathcal{V}(\mathbf{x}-\mathbf{y})\hat{\mu}(\mathbf{x})\mu(\mathbf{y}), \label{TDpot}
\end{equation}
where $\mu(\mathbf{y})$ is the measurement record of the mass density associated to the measurement process in Eq.~\eqref{TDmeas}:
\begin{equation}
\mu(\mathbf{x})=\ave{\hat{\mu}(\mathbf{x})}+\hbar \int \dd \mathbf{y}\, \gamma^{-1}(\mathbf{x}-\mathbf{y}) \delta \mu_t (\mathbf{y}). \label{massdenssign}
\end{equation}
Here, $\gamma^{-1}(\mathbf{x}-\mathbf{y})$ is {the inverse function of} $\gamma(\mathbf{x}-\mathbf{y})$, for which the following relation holds
\begin{equation}
(\gamma \circ \gamma^{-1})(\mathbf{x}-\mathbf{y})=\int \dd \mathbf{r} \,\gamma(\mathbf{x}-\mathbf{r})\gamma^{-1}(\mathbf{r}-\mathbf{y})=\delta(\mathbf{x}-\mathbf{y}). \label{gammainv}
\end{equation}
We report a method to construct the inverse kernel $\gamma^{-1}(\mathbf{x}-\mathbf{y})$ in Appendix \ref{Appendix C}.
The corresponding feedback wavefunction dynamics is {given by \cite{Tilloy2016}}
\bqali
{(\dd \ket{\psi})_\text{fb}}=&- \int \dd \x \dd \y \left\{  {\frac{i}{\hbar}\mathcal{V}(\x-\y)\mu(\y)} \right.\\
&\left. +\frac{1}{2}(\mathcal{V} \circ \gamma^{-1} \circ \mathcal{V})(\x-\y)\hat{\mu}(\y) \right\} \hat{\mu}(\x) \dd t {\ket{\psi}}. \label{TDfeed}
\eqali
We note that Eq.~\eqref{TDpot} and Eq.~\eqref{massdenssign} indicate that there is one measurement record at each point of space. Moreover, each constituent receives the same information about the mass density at a given point. Therefore, the TD model implements a universal interaction. See Fig.~\ref{TDgraph} for a graphical representation of the model.

The full evolution of the wavefunction is obtained by  merging Eq.~\eqref{TDmeas} and Eq.~\eqref{TDfeed}, yielding
\bqali
&\dd \ket{\psi}=\left({-\frac{i}{\hbar}\int \dd \mathbf{x} \dd \mathbf{y} \mathcal{V}(\mathbf{x}-\mathbf{y})\hat{\mu}(\mathbf{x}) \mu(\mathbf{y})\dd t} \right. \\
&-\frac{1}{2}\int \dd \mathbf{x} \dd \mathbf{y} \left(\mathcal{V} \circ \gamma^{-1} \circ \mathcal{V} \right)(\mathbf{x}-\mathbf{y})\hat{\mu}(\mathbf{x})\hat{\mu}(\mathbf{y}) \dd t \\
&-\frac{1}{8 \hbar^2}\int \dd \mathbf{x} \dd \mathbf{y} \gamma(\mathbf{x}-\mathbf{y}) \left( \hat{\mu}(\mathbf{x})-\ave{\hat{\mu}(\mathbf{x})} \right)\left(\hat{\mu}(\mathbf{y})-\ave{\hat{\mu}(\mathbf{y})} \right) \dd t\\
&+\frac{1}{2 \hbar} \int \dd \mathbf{x} \left(\hat{\mu}(\mathbf{x})-\ave{\hat{\mu}(\mathbf{x})} \right)\delta \mu_t (\mathbf{x}) \dd t \\
&\left.-\frac{i}{2 \hbar} \int \dd \mathbf{x} \dd \mathbf{y} {\mathcal{V}(\mathbf{x}-\mathbf{y})}\hat{\mu}(\mathbf{x})\left(\hat{\mu}(\mathbf{y})-\ave{\hat{\mu}(\mathbf{y})} \right)\dd t \right)\ket{\psi}.
\eqali
As in Eq.~\eqref{KTMfull}, such a dynamical equation now includes the feedback and the continuous measurement, as well as the It\^o term resulting from the combination of the two steps. The corresponding master equation reads
\bqali
\frac{\dd \hat{\rho}}{\dd t}=&-\frac{i}{\hbar}\left[\hat{H}_0+\hat{H}_{\text{grav}},\hat{\rho} \right]\\
&-\int \dd \mathbf{x} \dd \mathbf{y} D(\mathbf{x}-\mathbf{y})\left[\hat{\mu}(\mathbf{x}),\left[\hat{\mu}(\mathbf{y}),\hat{\rho}\right]\right], \label{TDmastereq}
\eqali
where $\hat{H}_{\text{grav}}$ is the Newtonian gravitational interaction defined in Eq.~\eqref{Hgrav};  we added the free Hamiltonian $\hat H_0$ and defined
\begin{equation}
D(\mathbf{x}-\mathbf{y})=\left[\frac{\gamma}{8 \hbar^2}+\frac{1}{2}\left(\mathcal{V} \circ \gamma^{-1} \circ \mathcal{V} \right)\right](\mathbf{x}-\mathbf{y}), \label{decohker}
\end{equation}
which is the decoherence kernel of the model. {The latter has a structure which is similar to that of the KTM master equation \eqref{KTMrecov}: one term is proportional to $\gamma$ while the second is inversely proportional to it. This shows the presence of a minimum, which can be retrieved by setting $\gamma(\x-\y)=-2 \hbar \mathcal{V}(\x-\y)$ \cite{Tilloy2017}. Such a correlation kernel leads to the decoherence rate of the Di\'osi-Penrose model \cite{Tilloy2017,Diosi1989,Penrose1996}.}
Similarly 
to the KTM model, TD model retrieves the quantum  gravitational interaction, whose unitary evolution is modified by the decoherence due to the measurement and the feedback dynamics.
The advantages of the TD model over the KTM model are two. First, one considers the full form of the Newtonian potential and not only its Taylor expansion near an equilibrium position. Second, the use of mass density operator allows to study also identical particles. 
In Appendix~\ref{SectTD} we discuss in detail the issues of divergences in the TD model and how to regularize them {through the use of a suitable smearing function $g(\x)$}.

\subsection{The KTM2 model}
As we will show in this subsection, a particular case of the TD model is given by the model described in Ref.~\cite{Kafri2015}, where the specific case of $N$ particles on a lattice is considered. We will refer to it as the KTM2 model, in order to avoid confusion with the model in Ref.~\cite{Kafri2014}. 

In this case the mass density operator reads ${\hat \mu(\x)=m\sum_\alpha \hat n_\alpha \delta (\x-\x_\alpha)}$, where $\hat n_\alpha$ is the local number density of the $\alpha$-th lattice site located at  position $\x_\alpha$. {Given the form of $\hat \mu(\x)$,  Eq.~\eqref{TDmastereq} becomes
\bqali\label{KTM2master}
\frac{\dd \hat{\rho}}{\dd t}=&-\frac{i}{\hbar}\left[\hat{H}_0+\hat{H}_{\text{grav}},\hat{\rho} \right]\\
&-\sum_{\alpha,\beta=1}^N m^2 D(\x_\alpha-\x_\beta) \left[\hat{n}_\alpha,\left[\hat{n}_\beta,\hat{\rho}\right]\right] ,
\eqali
where
\begin{equation}
\hat{H}_{\text{grav}}=\frac{m^2}{2}\sum_{\alpha,\beta=1}^N \mathcal{V}(\mathbf{x}_\alpha-\mathbf{x}_\beta)\hat n_\alpha\hat n_\beta. 
\end{equation} 
To avoid  divergences due to the self-interaction, one can regularize $\mathcal V$ with a suitable smearing function. The choice considered in Ref.~\cite{Kafri2015} is such that $m^2\mathcal{V}(\mathbf{x}_\alpha-\mathbf{x}_\beta)\to \chi_{\alpha \beta}=-Gm^2/[2(|\mathbf{x}
_\alpha-\mathbf{x}_\beta|+a)]$ where $a$ denotes a minimum length cutoff.
Now, by considering $\gamma(\x-\y)$ and $\gamma^{-1}(\x-\y)$ as non-negligible only for $\x-\y$ smaller than the lattice distance and considering that in such a case they read $\gamma(\x-\y)=2\hbar /m$ and $\gamma^{-1}(\x-\y)= m/2\hbar$, then  Eq.~\eqref{KTM2master} reduces to} 
\bqali
\frac{\dd \hat{\rho}}{\dd t}=&-\frac{i}{\hbar}\left[\hat{H}_0+\sum_{\alpha,\beta=1}^N \hat{V}_{\alpha \beta}, \hat{\rho}\right]- \frac{\xi}{2}\sum_{\alpha=1}^N \left[\hat{n}_\alpha,\left[\hat{n}_\alpha,\hat{\rho}\right]\right]\\
&-\frac{1}{2 \xi}\sum_{\alpha, \beta,\epsilon=1 }^N \chi_{\alpha \beta}\chi_{\alpha \epsilon}\left[\hat{n}_\beta,\left[\hat{n}_\epsilon,\hat{\rho}\right]\right],\label{KTM2}
\eqali
{where $\xi=m/2\hbar$. Equation \eqref{KTM2} coincides with the KTM2 master equation \cite{Kafri2015} once the self-interacting terms, although not being divergent, are removed by hand.}
 
\section{Comparison between the TD and KTM model}\label{Comparison}

The TD and KTM models consider the same problem: how to effectively implement the Newtonian gravitational interaction among  quantum systems by using a continuous measurement and a feedback. The way this is done is different in the two cases. 

In this section, we  compare the two models.  We first expand the gravitational interaction in the TD model to linear order. Then, by comparing the resulting master equations, we will see that the KTM model does not coincide with the linearized TD model.

Let us rewrite the position operator of each particle as follows
\begin{equation}
\hat{\x}_\alpha=\x^{(0)}_\alpha + \Delta \hat{\x}_\alpha,
\end{equation}
where $\Delta \hat{\x}_\alpha$ is the quantum displacement from a given position $\x^{(0)}_\alpha$.  For small displacements, we can approximate Eq.~\eqref{TDmastereq} as 
\bqali
\frac{\dd \hat{\rho}}{\dd t}\!=\!&-\!\frac{i}{\hbar}\left[\hat{H}_0,\hat{\rho}\right]\!+\!\frac{2 i \pi G}{\hbar}\!\!\sum_{\substack{\alpha, \beta=1 \\ \beta \neq \alpha}}^N \sum_{l,j=1}^3m_\alpha m_\beta \eta_{\alpha \beta 2 lj}\left[ \hat{x}_{\alpha l} \hat{x}_{\beta j},\hat{\rho}\right]\\
&-\sum_{\alpha,\beta=1}^N \sum_{l,j=1}^3 m_\alpha m_\beta \eta_{\alpha \beta lj}\left[\hat{x}_{\alpha l},\left[\hat{x}_{\beta j},\hat{\rho}\right]\right],\label{linearTD}
\eqali
where $\hat{x}_{\alpha l}$ is the component in the $l$ direction of $\Delta\hat\x_\alpha$. This choice of notation matches that used in Section~\ref{SectKTM}. 
Moreover, we included the terms coming from $\hat{H}_{\text{grav}}$ corresponding to the same particle ($\alpha=\beta$) in the definition of $\hat{H}_0$. {The parameter $\eta_{\alpha \beta lj} $ is defined as
\begin{equation}
\eta_{\alpha \beta lj}=\left(\frac{\pi^3}{8 \hbar^5}\right)^{1/2}\eta_{\alpha \beta 0 lj}+(8 \pi \hbar)^{1/2}G^2 \eta_{\alpha \beta 4 lj},
\end{equation}
and coefficients $\eta_{\alpha \beta n lj}$ are given by}
\bqali
\eta_{\alpha \beta 0 lj}&=\int \dd \k \,\tilde{g}^2(\k)\, \tilde \gamma(\k) k_l k_j\, {e^{-\frac{i}{\hbar} \k \cdot (\x^{(0)}_\alpha-\x^{(0)}_\beta )}},\\
\eta_{\alpha \beta 2 lj}&=\int \frac{\dd \k}{k^2}\,\tilde{g}^2(\k) k_l k_j \,e^{-\frac{i}{\hbar}\k \cdot (\x^{(0)}_\alpha-\x^{(0)}_\beta)},\\
\eta_{\alpha \beta 4 lj}&=\int \frac{\dd \k}{k^4}\,\frac{\tilde{g}^2(\k) }{\tilde{\gamma}(\k)}k_l k_j \,e^{-\frac{i}{\hbar}\k \cdot (\x^{(0)}_\alpha-\x^{(0)}_\beta )}, \label{coeffLinTD}
\eqali
{which are fixed once the correlation kernel $\gamma(\x-\y)$ and the Fourier transform $\tilde g(\k)$ of the smearing function $g(\x)$ are chosen. For example, one can take the Di\'osi-Penrose choice  $\gamma(\x-\y)=-2 \hbar \mathcal{V}(\x-\y)$ and a Gaussian smearing function. }

In the case of two particles in one dimension, Eq.~\eqref{linearTD} reduces to: 
\bqali
\frac{\dd \hat{\rho}}{\dd t}\!=\!&-\!\frac{i}{\hbar}\left[\hat{H}_0 + \hat{H}_{\text{grav}}
,\hat{\rho}\right]
-\sum_{\alpha=1}^2  m_\alpha^2 \eta_{\alpha \alpha }\left[\hat{x}_{\alpha },\left[\hat{x}_{\alpha },\hat{\rho}\right]\right]\\
&-  m_1 m_2 \left(\eta_{1 2 }\left[\hat{x}_{1 },\left[\hat{x}_{2 },\hat{\rho}\right]\right] +\eta_{2 1 }\left[\hat{x}_{2 },\left[\hat{x}_{1 },\hat{\rho}\right]\right]
\right),
\label{TD2}
\eqali
which clearly differs from Eq.~\eqref{KTMrecov}. While the double commutator term in Eq.~\eqref{KTMrecov} contains only the position operators corresponding to the same particle, the corresponding term in Eq.~\eqref{TD2} contains also position operators of different particles. The result of Eq.~\eqref{linearTD} differs from both the pairwise and universal generalisations of the KTM model [cf.~Eq.~\eqref{Altamirano} and Eq.~\eqref{KTMfake}] for the same reason. {Therefore, the TD model cannot be reduced to that of KTM, or, \textit{viceversa}, the TD model is not a generalization of the KTM model to continuous mass densities and full gravitational interactions.}

The fact that the two models are different is not surprising, because they implement the measurement and feedback mechanism in two different ways. In the KTM model and its generalizations considered in Section \ref{sec.3}, one or more noises are attached to the position of the particle and they follow it while it moves in space. In the TD model, there is one noise for each point of space;  these noises do not follow the particle, rather the particle feels different noises while moving in space. This is the ultimate reason for the difference between the two models.   

{Finally, we underline that Eq.~\eqref{linearTD} does not suffer from the limitations of the generalizations of the KTM model in Eq.~\eqref{Altamirano} and Eq.~\eqref{KTMfake}. On one hand, the TD model is implemented through an universal measurement and feedback protocol -- a single measurement is performed -- instead of a pairwise one, where $N-1$ measurements take place for each particle. Thus, the decoherence effects do not scale with the number of measurements $N-1$ as in Eq.~\eqref{Altamirano}. On the other hand, the TD model is self-consistent: when considering the system as divided in two subsystems of $N_1$ and $N_2$ particles respectively, the master equation describing the center-of-mass motion of these subsystems coincides with that of two particles given by $N=2$, and in both master equations there will be present double-commutator terms containing operators of different particles, see for instance the linearized case in Eq.~\eqref{TD2}. This is simply a feature of the model, which differs from that of KTM. This is even more explicit  by comparing the KTM master equation \eqref{KTMrecov} and that in Eq.~\eqref{TD2} for the TD model in the linear case for $N=2$.}
We also remind that the TD model easily accounts for identical particles by properly writing the mass-density operator in a quantum field theoretical language. On the contrary, it is not obvious to see how this can be achieved in the KTM model.

\section{Discussion and conclusions}

The main virtue of the KTM \cite{Kafri2014} and TD \cite{Tilloy2017} models is that the {Newtonian} gravitational interaction is  implemented within a hybrid classical-quantum framework, where gravity is classical and matter is quantum, thus showing that this possibility is not inconsistent. The price to pay are additional decoherence effects which can be minimized but not fully evaded. 

In this work, we discussed the generalization of the KTM model to $N$ particles, keeping the original spirit of a continuous measurement of the position of the particles and subsequent feedback evolution, which together reproduce the linearized Newtonian potential. The pairwise generalization was shown to be incompatible with experimental data \cite{Altamirano2018}, whereas the universal one is inconsistent. Therefore the two most natural ways to generalize the KTM model are not viable.

Next, we considered the TD model \cite{Tilloy2017} model and we showed that, when reduced to two particles in one dimension, in the limit of a linearized Newtonian potential, it does not reproduce the KTM model, which then is not an approximated version of the TD model.

The KTM and TD models implement the continuous measurement plus feedback protocol in two different ways. In the first case the position of the particle is measured, in the second case the mass density is measured. This is the difference, which ultimately allows TD to consider the full Newtonian potential, not only its linearized limit as for the KTM model. In a nutshell, the reason is that the mass of a particle enters the Newtonian potential linearly,  therefore the standard theory of linear feedback can be used: see Section~\ref{TDmodsect}. The position instead enters nonlinearly (at the denominator), and the standard formalism cannot be applied any longer \cite{Diosi:2017aa}. In fact, suppose that the position of the particle is measured and one uses it to write the feedback Hamiltonian, in analogy with what discussed in the previous sections. This would look like 
\begin{equation}
\hat{H}_{\text{fb}}=-\sum_{\substack{\alpha, \beta=1 \\ \beta \neq \alpha}}^N \frac{G m_\alpha m_\beta}{|\hat{\x}_\alpha -\r_\beta|},
\end{equation}
where the measurement record $\r_\alpha$ enters nonlinearly. 
This  nonlinearity  does not allow to implement the prescription of Eq.~\eqref{feedbackS} to obtain the feedback contribution to the dynamics.

Although it is `morally' the same to measure the position of the particles or their mass density, these two are different operations. 
When measuring the position, the noise is attached to the particle and follows its position. Conversely, when measuring the mass density, there is a noise for each point of space: if the particle moves, different noises act on it. One consequence of this difference is that the resulting master equations are different, because they correspond to two different measurement schemes. This is why the KTM and TD models, when compared in the same regime of applicability, give different results. 

A natural open question is whether this approach can be generalized to a relativistic setting. This will be subject of future research.

\section*{Acknowledgements}
J.L.G.R. thanks L. Asprea, F. Benatti, G. Gasbarri, A. Gundhi, and C. Jones for the helpful discussions and comments. M.C. and A.B. acknowledge financial support from the H2020 FET Project TEQ (Grant No. 766900) and the support by grant number (FQXi-RFP-CPW-2002) from the Foundational Questions Institute and Fetzer Franklin Fund,  a donor advised fund of Silicon Valley Community Foundation. J.L.G.R. acknowledges financial support from The Abdus Salam ICTP. A.B. acknowledges financial support from the COST Action QTSpace (CA15220), INFN and the University of Trieste.

\appendix

\section{Continuous measurements  and Feedback} \label{AppendixA}

We recall here the main properties characterizing the continuous measurement of a  Hermitian operator $\hat{a}$. The results follow mainly from Refs.~\cite{Jacobs2006, Jacobs2014}. 

We consider a continuous observable $\hat{a}$ with associated eigenstates $\left\{\ket{a} \right\}_{a \in \mathbb{R}}$ satisfying $\hat{a}\ket{a}=a\ket{a}$. One divides time into (infinitesimal) intervals of length $\Delta t$. In each interval, one makes the weak measurement described by the operator
\begin{equation}
\hat{A}(r)=\left(\frac{\gamma \Delta t}{2 \pi \hbar^2} \right)^{1/4} \int_{-\infty}^{\infty} \dd a\,\exp \left[-\frac{\gamma \Delta t}{4 \hbar^2}  \left(a -r \right)^2 \right] \ket{a}\bra{a}. \label{measuroper}
\end{equation}
One then obtains a continuum of measurement results labelled by this parameter $r$. Denoting by $P(r) = \langle\psi| \hat{A}^{\dagger}(r)\hat{A}(r) | \psi\rangle$ the probability density of the measurement result $r$, the mean value $\ave{r}$ of $r$, and the variance $\sigma_r^2$ of $r$ are related to those of $\hat{a}$ by
\begin{equation}
\ave{r}\!=\!\int_{-\infty}^{\infty} r P(r) \dd r \!=\!\ave{\hat{a}}, \quad \sigma_r^2\!=\!\ave{r^2}-\ave{r}^2\!=\!\sigma_{\hat{a}}^2 + \frac{\hbar^2}{\gamma \Delta t}. \label{meanvariation}
\end{equation}
Since the time interval $\Delta t$ is infinitesimal, the probability density $P(r)$ can be approximated as
\begin{equation}
P(r) \approx \frac{1}{\hbar}\sqrt{\frac{\gamma \Delta t}{2 \pi}} \exp \left[-\frac{\gamma \Delta t}{2 \hbar^2} \left(r-\ave{\hat{a}}\right)^2 \right]. \label{probdensity}
\end{equation}
From the results of Eq.~\eqref{meanvariation} and Eq.~\eqref{probdensity}, $r$ can be written as a stochastic quantity
\begin{equation}
r=\ave{\hat{a}}+\frac{\hbar}{\sqrt{\gamma}}\frac{\Delta W_t}{\Delta t},
\end{equation}
where $\Delta W_t$ is  a Gaussian random variable with zero mean and variance $\Delta t$. 

By performing a sequence of these measurements, and taking the limit $\Delta t \rightarrow 0$, one obtains a so-called continuous measurement, described by
\begin{equation}
r=\ave{\hat{a}}+\frac{\hbar}{\sqrt{\gamma}}\frac{\dd W_t}{\dd t}. \label{contmesrec}
\end{equation}
In the above equation, the parameter $\gamma$ is the information rate gained by the measurement, and $W_t$ is a standard Wiener process, satisfying
\begin{equation}
\mathbb{E}[\dd W_t]=0, \qquad \mathbb{E}[(\dd W_t)^2]=\dd t
\end{equation}
We can see that the measurement records defined in Eq.~\eqref{KTMsign} are a specific application of Eq.~\eqref{contmesrec} where the observables measured are the position operators $\hat{x}_\alpha$ of the particles, with $\alpha=1,2$. 

Let us denote by $\ket{\psi}$ the state of a system at a time $t$ before performing a continuous measurement of the observable $\hat{a}$. The evolution of the system will be described by applying the operator $\hat{A}(r)$ to the state $\ket{\psi}$, and performing the limit $\Delta t \rightarrow 0$. By demanding that the resulting dynamical equation preserves the norm, one obtains
\begin{equation}
{(\dd \ket{\psi})_{\text{m}}} \!=\! \left\{-\frac{\gamma}{8 \hbar^2}\left(\hat{a}-\ave{\hat{a}} \right)^2 \dd t + \frac{\sqrt{\gamma}}{2 \hbar} \left(\hat{a}-\ave{\hat{a}} \right) \dd W_t \right\} {\ket{\psi}}, \label{measformalism}
\end{equation}
so that the result of Eq.~\eqref{measurement} is consistent with the general formalism of Eq.~\eqref{measformalism}. The generalization to a continuous set of observables used in Section~\ref{TDmodsect} can be found in Ref.~\cite{Tilloy2016}.

Quantum feedback is implemented to modify and control the evolution of a system \cite{Jacobs2014}. In this Appendix, we review the derivation of the Wiseman-Milburn Markovian feedback master equation \cite{Wiseman1993, Wiseman2010, Zhang2017}. The derivation follows the approach of Ref.~\cite{Diosi}.
 
In the Markovian case, the feedback Hamiltonian $\hat{H}_{\text{fb}}$ is expressed in terms of the measurement record $r$ of the observable $\hat{a}$ as 
\begin{equation}
\hat{H}_{\text{fb}}=r \hat{b}, \label{Markovianfeed}
\end{equation}
where $\hat{b}$ is a Hermitian operator. The feedback evolution can be obtained by unitarily evolving the state of the system $\ket{\psi}$ \cite{Diosi}. This  gives
\begin{equation}
e^{-\frac{i}{\hbar}\hat{H}_{\text{fb}}\dd t}\ket{\psi} = \ket{\psi}+ {(\dd \ket{\psi})_{\text{fb}}},
\end{equation}
where ${(\dd \ket{\psi})_{\text{fb}}}$ turns out to be:
\begin{equation}
{ (\dd \ket{\psi})_{\text{fb}}}=\left(\left[-\frac{i}{\hbar}\ave{\hat{a}}\hat{b}-\frac{1}{2 \gamma}\hat{b}^2 \right] \dd t -\frac{i}{\sqrt{\gamma}}\hat{b} \, \dd W_t \right) {\ket{\psi}}. \label{feedbackformal}
\end{equation}

The combined measurement + feedback evolution of the system is obtained by considering the contributions of both the continuous measurement of $\hat{a}$ [{cf.}~Eq.~\eqref{measformalism}]  and the subsequent feedback dynamics driven by $\hat{b}$ as described by Eq.~\eqref{feedbackformal} \cite{Jacobs2014,Diosi}. In an infinitesimal time $\dd t$, the wavefunction of the system is given by $\ket{\psi(t+\dd t)}=\ket{\psi}+ \dd \ket{\psi}$, where
\begin{equation}
\dd \ket{\psi}= (\dd \ket{\psi})_{\text{m}}+ (\dd \ket{\psi})_{\text{fb}}+ {(\dd \ket{\psi})_{\text{m+fb}}}
\label{wavefundis}
\end{equation}
The first two terms of the stochastic differential equation for the wavefunction are given by Eq.~\eqref{measformalism} and Eq.~\eqref{feedbackformal}, while the contribution in the last term comes {from the product of the noise terms in the differential equations for the measurement and  the feedback,} {i.e. from the application of $e^{i \hat H_\text{fb}\dd t/\hbar}$ to the post-measurement state $\ket \psi+(\dd \ket{\psi})_{\text{m}}$ approximated to the first order in $\dd t$}. This term is explicitly given by
\begin{equation}
{(\dd \ket{\psi})_\text{m+fb}}=-\frac{i}{2 \hbar}\hat{b} \left( \hat{a}-\ave{\hat{a}} \right) \dd t \ket{\psi}.
\end{equation}

From the definition of the density operator in terms of the wavefunction, $ \hat{\rho}=\mathbb{E}\left[\ket{\psi}\bra{\psi} \right]$, it follows that 
\bqali
\dd \hat{\rho} &=\dd (\mathbb{E}[\ket{\psi}\bra{\psi}])\\
 &= \mathbb{E}[ (\dd \ket{\psi})\bra{\psi}+ \ket{\psi} (\dd {\bra{\psi}})+ (\dd \ket{\psi})(\dd \bra{\psi})].\label{stochdens}
\eqali
Therefore, we can derive the master equation, by using Eq.~\eqref{wavefundis}. One obtains
\begin{equation}
\frac{\dd \hat{\rho}}{\dd t}=-\frac{i}{2 \hbar}[\hat{b},\left\{\hat{a},\hat{\rho} \right\}]-\frac{\gamma}{8 \hbar^2}\left[\hat{a},\left[\hat{a},\hat{\rho}\right]\right]-\frac{1}{2 \gamma}[\hat{b},[\hat{b},\hat{\rho}]].
\end{equation}

{We now generalize the procedure to $\mathcal{M}$ measurements. Consider a set of observables with associated Hermitian operators $\left\{\hat{a}_\alpha \right\}_{\lambda=1}^\mathcal{M}$, which are continuously measured. The corresponding measurement records read
\begin{equation}
r_\lambda=\ave{\hat{a}_\lambda}+\frac{\hbar}{\sqrt{\gamma_\lambda}}\frac{\dd W_{\lambda,t}}{\dd t},
\end{equation}
with $\gamma_\lambda$ denoting the information rates and $W_{\lambda,t}$ standard independent Wiener processes, satisfying
{
\bqali
\mathbb{E}[\dd W_{\lambda,t}] &=0\\
\mathbb E\left[\dd W_{\lambda,t} \dd  W_{\lambda',t}\right]&=\delta_{\lambda \lambda'}\dd t,
\eqali}
The stochastic differential equation for the continuous measurement is given by the sum of all the contributions due to each measurement, i.e.
\bqali
(\dd \ket{\psi})_\text{m}=\sum_{\lambda=1}^\mathcal{M}&\left\{- \frac{\gamma_\lambda}{8 \hbar^2}(\hat{a}_\lambda-\ave{\hat{a}_\lambda})^2 \dd t \right.\\
&\left.+  \frac{\sqrt{\gamma_\lambda}}{2 \hbar}(\hat{a}_\lambda-\ave{\hat{a}_\lambda})\dd W_{\lambda,t}\right\}\ket{\psi}, \label{continuousmeasurementseveral}
\eqali
and for a feedback Hamiltonian $\hat{H}_{\text{fb}}$ of the form
\begin{equation}
\hat{H}_{\text{fb}}=\sum_{\lambda=1}^\mathcal{M} r_\lambda \hat{b}_\lambda, \label{feedbackmultiple}
\end{equation}
with $\left\{ \hat{b}_\lambda \right\}_{\lambda=1}^\mathcal{M}$ a set of Hermitian operators, we obtain
\begin{equation}
(\dd \ket{\psi})_{\text{fb}}\!=\!\!\sum_{\lambda=1}^\mathcal{M} \!\! \left\{\!\!\left[\!-\frac{i}{\hbar}\ave{\hat{a}_\lambda} \hat{b}_\lambda \!\!-\! \!\frac{1}{2 \gamma_\lambda}\hat{b}_\lambda^2 \right]\!\dd t \!-\!\frac{i}{\sqrt{\gamma_\lambda}}\hat{b}_\lambda \dd W_{\lambda,t} \! \right\} \! \!\ket{\psi}.
\end{equation}
The stochastic differential equation for the wavefunction is given by Eq.~\eqref{wavefundis}, where now
\begin{equation}
(\dd \ket{\psi})_{\text{fb}}(\dd \ket{\psi})_{\text{m}}=-\frac{i}{2\hbar}\sum_{\lambda=1}^\mathcal{M} \hat{b}_\lambda (\hat{a}_\lambda-\ave{\hat{a}_\lambda}) \dd t.
\end{equation}
From Eq.~\eqref{stochdens}, the density operator satisfies
\bqali
\frac{\dd \hat{\rho}}{\dd t}=\sum_{\lambda=1}^\mathcal{M}&\left(-\frac{i}{2 \hbar}\left[\hat{b}_\lambda,\left\{\hat{a}_\lambda,\hat{\rho}\right\}\right]-\frac{\gamma_\lambda}{8 \hbar^2}\left[\hat{a}_\lambda,\left[\hat{a}_\lambda,\hat{\rho}\right]\right]\right.\\
&\left.-\frac{1}{2 \gamma_\lambda}\left[\hat{b}_\lambda,\left[\hat{b}_\lambda,\hat{\rho}\right]\right] \right). \label{MEcontfeedMop}
\eqali
From Eq.~\eqref{MEcontfeedMop}, one can derive the master equations of the two generalizations of the KTM model corresponding to Eq.~\eqref{Altamirano} and Eq.~\eqref{KTMfake}.
In particular, the master equation implementing the measurement and feedback through a pairwise protocol, i.e.~Eq.~\eqref{Altamirano}, is obtained by using $\mathcal M=9N(N-1)$ measurement records $\set{r_\lambda}_\lambda$, which are identified by four indices: $l$ and $j$ run over the three Cartesian directions, $\alpha$ identifies one among the $N$ measured particle and $\beta\neq\alpha$ identifies one among the remaining $N-1$ particles to which the information is sent. Namely, one imposes
\bqali
&\set{\hat a_\lambda}_\lambda\to\set{\hat x_{\alpha l}}_{\alpha l}, \quad\forall \beta, j\\
&\set{\hat b_\lambda}_\lambda\to \set{K_{\alpha \beta lj}\hat{x}_{\beta j}}_{\alpha \beta lj}, \\
&\set{\gamma_\lambda}_\lambda\to\set{\gamma_{\alpha \beta lj}}_{\alpha \beta lj},
\eqali
in Eq.~\eqref{MEcontfeedMop} and obtains Eq.~\eqref{Altamirano}.\\
The universal generalization of the KTM model, i.e.~Eq.~\eqref{KTMfake}, is instead easily provided by imposing 
\bqali
&\set{\hat a_\lambda}_\lambda\to\set{\hat x_{\alpha l}}_{\alpha l},\\
&\set{\hat b_\lambda}_\lambda\to\{\sum_{\substack {\beta=1 \\ \beta \neq \alpha}}^N\sum_{j=1}^3 \chi_{\alpha \beta lj} \hat{x}_{\beta j}\}_{\alpha l},\\
&\set{\gamma_\lambda}_\lambda\to\set{\gamma_{\alpha l}}_{\alpha l},
\eqali
in Eq.~\eqref{MEcontfeedMop} with $\mathcal M=3N$.

\section{Construction of the correlation kernels} \label{Appendix C}

We describe with more detail the relation between a kernel $\mathcal{K}(\mathbf{x}-\mathbf{y})$ and its inverse $\mathcal{K}^{-1}(\mathbf{x}-\mathbf{y})$ by following the approach developed in Ref.~\cite{Ulmer2003}. Consider the operator $\mathcal{A}$ which satisfies 
\begin{equation}
\mathcal{A}\mathcal{K}(\mathbf{x}-\mathbf{y}) = \delta (\mathbf{x}-\mathbf{y}), \label{kernelgreen}
\end{equation}
where $\mathcal{K}(\mathbf{x}-\mathbf{y})$ is the associated kernel. We define the integral transform
\begin{equation}
u(\mathbf{x})=\int  \dd \mathbf{r} \mathcal{K}(\mathbf{r}-\mathbf{x})f(\mathbf{r}), \label{utransform}
\end{equation}
and require that the inverse kernel $\mathcal{K}^{-1}(\x-\y)$ satisfies
\begin{equation}
\delta(\mathbf{x}-\mathbf{y})=\int \dd \mathbf{r} \mathcal{K}(\mathbf{x}-\mathbf{r})\mathcal{K}^{-1}(\mathbf{r}-\mathbf{y}).
\end{equation}
From these expressions, we can show that
\begin{equation}
f(\x)=\int \dd \r \mathcal{A} \mathcal{K}(\r-\x)f(\r), \label{fK}
\end{equation}
and equivalently
\begin{equation}
f(\x)=\int \dd \r \mathcal{K}^{-1}(\r-\x)u(\r). \label{fKinv}
\end{equation}
The substitution of Eq.~\eqref{utransform} in Eq.~\eqref{fKinv} and the comparison with Eq.~\eqref{fK} lead to
\begin{equation}
\mathcal{K}^{-1}(\mathbf{x}-\mathbf{y})=\mathcal{A}^2 \mathcal{K}(\mathbf{x}-\mathbf{y})=\mathcal{A}\delta(\mathbf{x}-\mathbf{y}), \label{invkernel}
\end{equation}
where the last equality follows from Eq.~\eqref{kernelgreen}. In the following we consider two examples. First, let us take 
\begin{equation}
\mathcal{A}=\frac{1}{4 \pi G}\nabla^2, \qquad \mathcal{K}(\mathbf{x}-\mathbf{y})=-\frac{G}{|\mathbf{x}-\mathbf{y}|},
\end{equation}
then from Eq.~\eqref{invkernel}, we have
\begin{equation}
\mathcal{K}^{-1}(\mathbf{x}-\mathbf{y})=\frac{1}{4 \pi G}\nabla^2 \delta (\mathbf{x}-\mathbf{y}).
\end{equation}
A less trivial example is that of the operator
\begin{equation}
\mathcal{A}=\exp \left[-\frac{1}{4} \sigma^2 \nabla^2 \right],
\end{equation}
and the kernel
\begin{equation} \mathcal{K}(\mathbf{x}-\mathbf{y})=\frac{1}{(\pi \sigma^2)^{3/2}}\exp \left[-\frac{(\mathbf{x}-\mathbf{y})^2}{\sigma^2} \right].
\end{equation}
Then, it can be shown \cite{Ulmer2003} that
\begin{equation}
\mathcal{K}^{-1}(\mathbf{x}-\mathbf{y})=\mathcal{K}(\mathbf{x}-\mathbf{y})\prod_{k=1}^3 \sum_{n_k=0}^\infty c_{n_k} H_{2n_k}\left(\frac{x_k-y_k}{\sigma} \right),
\end{equation}
where $H_{2n_k}$ are the Hermite polynomials of degree $2n_k$, and $c_{n_k}=(-1)^{2 n_k}/(2^{n_k}n_k!)$ \footnote{The reported coefficients in Ref.~\cite{Ulmer2003} are $c_{n_k}~=~(-1)^{2 n_k}\sigma^{2 n_k}/(2^{n_k}n_k!)$ and differ from our calculations.}. 

{\section{Inconsistency of the universal generalization of the KTM model}\label{coefUni}

We showed in the main text that, when reducing the master equation \eqref{KTMfake} of the universal generalization of the KTM for $N$ particles to that for the center-of-mass, additional terms appear with respect to the master equation of the original KTM model, and as such the universal KTM model is inconsistent. Here, we provide an explicit example proving that coefficient multiplying the double commutator in Eq.~\eqref{eq.strage.double} in general is non-vanishing. We take the case of $N=3$ masses aligned along one dimension, which are then aggregated as displayed in Figure~\ref{fig.example}. 
\begin{figure}[t]
\centering
\includegraphics[width=\linewidth]{{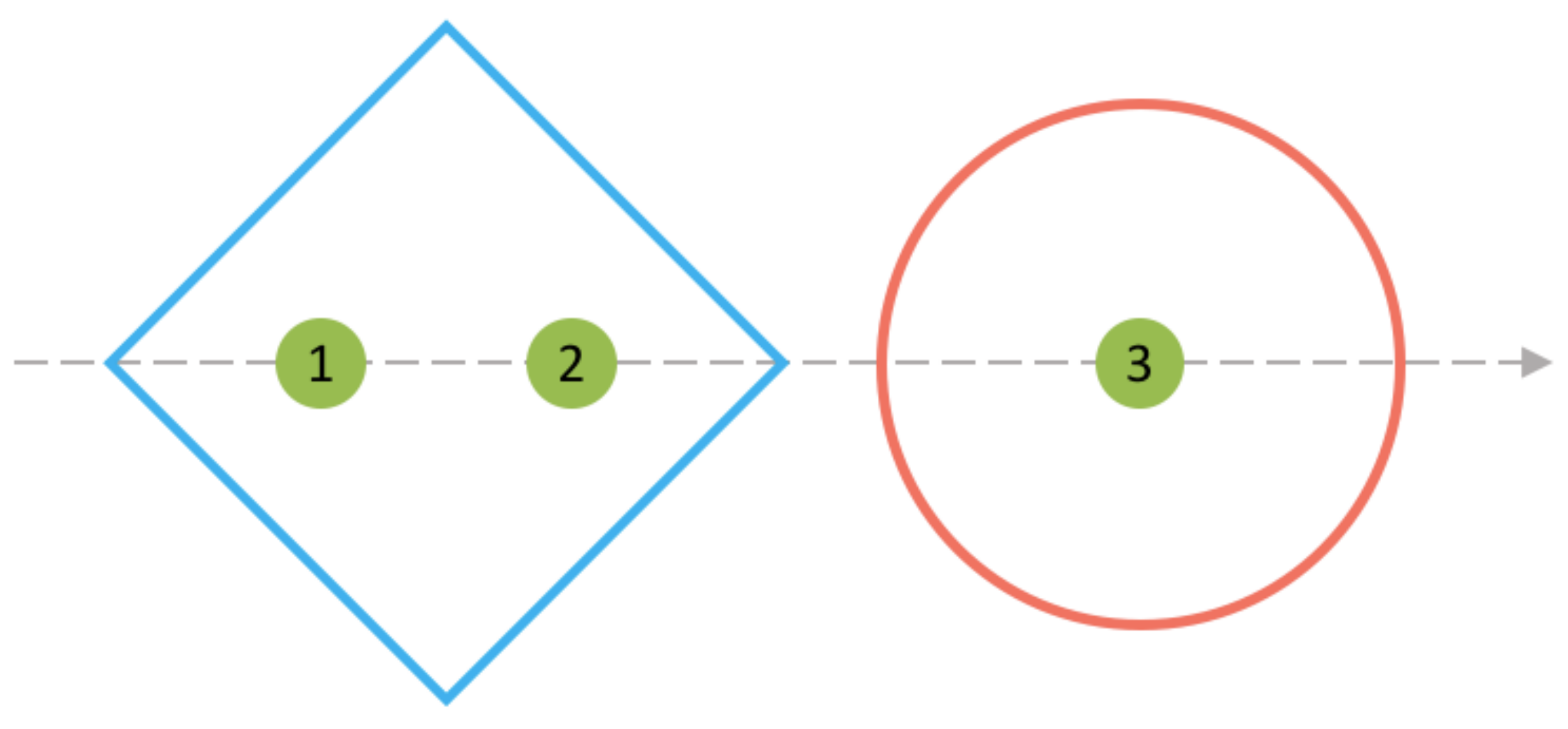}}
\caption{\label{fig.example} Example of lack of scale invariance in the universal KTM model with three masses, which are identified by the green spheres and are aligned along one dimension.  They are aggregated as two composite systems: the blue square $\diamond$ and the red circle $\bullet$.}
\end{figure}
Under such an assumption, the last term of Eq.~\eqref{KTMfake} becomes
\bq\label{eq.boublecheck}
-\sum_{\substack{\alpha,\beta, \epsilon=1 \\ \beta, \epsilon \neq \alpha}}^N\frac{K_{\alpha \beta }K_{\alpha \epsilon }}{2 \gamma_{\alpha }}\left[\hat{x}_{\beta },\left[\hat{x}_{\epsilon },\hat{\rho}\right]\right],
\eq
where
\bq
K_{\alpha\beta}=\frac{2Gm_\alpha m_\beta}{d_{\alpha\beta}^3},
\eq
is obtained from Eq.~\eqref{defK3d} by setting $l=j$ and $|{\bf d}_{\alpha\beta}|=d_{\alpha\beta}$. Note that $K_{\alpha\beta}>0$ for any value of $\alpha$ and $\beta$, which implies that also the coefficient in front of the double commutator in Eq.~\eqref{eq.boublecheck} is positive. Now, we express the position operators $\hat x_\alpha$ as sum of the center-of-mass position operator $\hat X_\alpha^\mu$ and relative position operator $\hat x_\alpha^\mu$. In particular, in accordance with the division displayed in Fig.~\ref{fig.example}, we have $\mu=\diamond$ for $\alpha=1$ or 2; while $\mu=\bullet$ for $\alpha=3$. Then, by tracing over the relative degrees of freedom [cf.~Eq.~\eqref{eq.strage.double}], Eq.~\eqref{eq.boublecheck} reduces to
\bqali
&-\mathcal S^{\diamond\diamond}\left[\hat X^\diamond,\left[\hat X^\diamond,\hat{\rho}_\text{\tiny CM}\right]\right]
-\mathcal S^{\bullet\bullet}\left[\hat X^\bullet,\left[\hat X^\bullet,\hat{\rho}_\text{\tiny CM}\right]\right]\\
&-\mathcal S^{\bullet\diamond}\left[\hat X^\bullet,\left[\hat X^\diamond,\hat{\rho}_\text{\tiny CM}\right]\right].
\eqali
The last term of this equation is the additional term with respect to the original KTM model. The explicit expression of its coefficient is 
\bq
\mathcal S^{\bullet\diamond}=2\left(\frac{K_{12}K_{13}}{\gamma_1}+\frac{K_{21}K_{23}}{\gamma_2}
\right),
\eq 
which is always strictly positive, as noted above. This proves that such additional terms in general are non vanishing.

}

\section{The divergences and regularization in the TD model} \label{SectTD}

The decoherence term~\eqref{decohker}  in the master equation \eqref{TDmastereq} is only formally defined.  We show that, under the assumption that $\gamma(\mathbf{x},\mathbf{y})$ is invariant under translations, i.e. $\gamma(\mathbf{x},\mathbf{y})=\gamma(\mathbf{x}-\mathbf{y})$, any choice of $\gamma$ leads to divergences. 
To do so, let us consider a system of point-like particles, whose mass density is given by
\begin{equation}
\hat{\mu}(\mathbf{x})=\sum_{\alpha=1}^N m_\alpha \delta (\mathbf{x}-\hat{\x}_\alpha). \label{pointlikemd}
\end{equation}
By substituting it, once expressed in terms of its Fourier transform, in the decoherence term of Eq.~\eqref{TDmastereq}, we obtain
\bqali
&\int \dd \x \dd \y D(\x,\y)\left[\hat{\mu}(\x),\left[\hat{\mu}(\y),\hat{\rho}\right]\right]\\
&{=\sum_{\alpha,\beta=1}^Nm_\alpha m_\beta \int \frac{\dd \k \tilde D(\k)}{(2\pi\hbar)^{3/2}}\left[e^{-\frac{i}{\hbar} \k\cdot\hat\x_\alpha},\left[e^{\frac{i}{\hbar} \k\cdot\hat\x_\beta},\hat{\rho}\right]\right]},\label{TDprob}
\eqali
where $\tilde D(\k)$ is the Fourier transform of $D(\x-\y)$, which inherits the translational invariance from $\gamma$. Let us consider the terms in the above sum corresponding to the same particle ($\alpha=\beta$). These are proportional to
\bq
\int\dd \k\,\tilde D(\k)\left(2\hat \rho-e^{-\frac{i}{\hbar} \k\cdot\hat\x_\alpha}\hat{\rho}e^{\frac{i}{\hbar} \k\cdot\hat\x_\alpha}-e^{\frac{i}{\hbar} \k\cdot\hat\x_\alpha}\hat{\rho}e^{-\frac{i}{\hbar} \k\cdot\hat\x_\alpha}\right).
\eq
The first term $\int \dd \k \tilde{D}(\k)$ diverges;
a straightforward calculation show that {according to Eq.~\eqref{decohker}}:
\begin{equation}
\int \dd \k \,\tilde{D}(\k)=\int \dd \k \left(\frac{\tilde{\gamma}(\k)}{8 \hbar^2}+8\pi^2\hbar^4G^2 \frac{\widetilde{\gamma^{-1}}(\k)}{k^4} \right), \label{decokerFour}
\end{equation}
where the Fourier transform of the inverse of the noise kernel $\widetilde{\gamma^{-1}}(\k)$ is related to $\tilde{\gamma}(\k)$ via
Eq.~\eqref{gammainv}:
\begin{equation}
\tilde{\gamma}(\k)\widetilde{\gamma^{-1}}(\k)=\frac{1}{(2 \pi \hbar)^3}, \label{gammainvFour}
\end{equation}
Then, Eq.~\eqref{decokerFour} can be written in terms of $\tilde{\gamma}(\k)$ as
\begin{equation}
\int \dd \k\,  \tilde{D}(\k)=\int \dd \k \left(\frac{\tilde{\gamma}(\k)}{8 \hbar^2}+\frac{\hbar G^2}{\pi}\frac{1}{k^4 \tilde{\gamma}(\k)} \right). \label{decoker2}
\end{equation}
Equation \eqref{decoker2} is the sum of two contributions: the continuous measurement, which gives the first term, and the application of the gravitational interaction through a feedback evolution, which provides the second term.
Before analysing the general case, let us study two particular correlation kernels. 

The first case corresponds to a LOCC dynamics, which requires that the dynamics acts only locally \cite{Tilloy2017}. A noise correlation function reflecting this property is proportional to a Dirac-delta. Thus, we set
\begin{equation}
\gamma(\x-\y)=A \delta (\x-\y), \label{corrdel}
\end{equation}
where $A$ is an arbitrary constant. In such a case, we have that $\tilde{\gamma}(\k)=A/(2 \pi \hbar)^{3/2}$. By substituting the latter expression in Eq.~\eqref{decoker2}, one gets that none of its contributions is convergent. 
Thus, in the TD model, the assumptions of having point-like particles and implementing a LOCC dynamics lead to divergences. 

As second case of interest, we consider a Gaussian correlation kernel $\gamma(\z)=(2 \pi \sigma^2)^{-3/2}\exp \left[-\z^2/(2 \sigma^2) \right]$. In this case, one has $\tilde{\gamma}(\k)=(2 \pi \hbar)^{-3/2}\exp(-\k^2 \sigma^2/2 \hbar^2) $. Now, by substituting the latter expression in Eq.~\eqref{decoker2}, we find that although the continuous measurement contribution converges, the feedback contribution is still divergent. 

Next, we show the general case: any choice of $\gamma(\x-\y)$ leads to divergences. Similarly to what was done in the KTM model, we minimize the decoherence kernel $\tilde{D}(\k)$ with respect to $\tilde{\gamma}(\k)$. The minimum is reached for $\tilde{\gamma}(\k)= G (2 \pi\hbar)^{3/2}/( \pi^2 k^2)$, which corresponds to $\gamma(\x-\y)=-2 \hbar \mathcal{V}(\x-\y)$. Such a correlation kernel leads to the decoherence rate of the Di\'osi-Penrose model \cite{Tilloy2017,Diosi1989,Penrose1996}, which is still divergent \cite{Bahrami2014}. Indeed, Eq.~\eqref{decoker2} in this case reads
\begin{equation}
\int \dd \k \tilde{D}(\k)=\frac{2(2\pi\hbar)^{1/2}G}{\hbar}\int_{0}^\infty \dd k \rightarrow \infty.
\end{equation}
Since the latter choice of $\gamma(\x-\y)$ provides the minimum decoherence rate, we deduce that Eq.~\eqref{decoker2}, and subsequently the master equation \eqref{TDmastereq}, diverges for any choice of $\gamma(\x-\y)$.\\

A regularization process is needed to avoid divergences in the decoherence terms in  Eq.~\eqref{TDmastereq}. This regularization mechanism is   applied also to the Di\'osi-Penrose model \cite{Bahrami2014}, by introducing a smearing function. 
For the TD model, the contributions to the decoherence term are those coming from the measurement part, through $\gamma(\x-\y)$, and from the feedback evolution, through $(\mathcal{V} \circ \gamma^{-1} \circ \mathcal{V}^{-1})(\x-\y)$. Both  these terms must be regularized. Indeed, the regularization of the noise kernel $\gamma(\x-\y)$ alone would only give a different noise kernel $\gamma'(\x-\y)$, which is not sufficient to avoid the divergence, as proved before. 
On the other hand, the regularization of the  gravitational potential $\mathcal{V}(\mathbf{x}-\mathbf{y})$ could remove the divergences in the feedback contribution, but not that due to the measurement, which is independent from the  gravitational interaction.  We conclude that the regularization mechanism must be performed by smearing both $\gamma(\x-\y)$ and $\mathcal{V}(\x-\y)$.

An effective regularization procedure consists {in} smearing the mass density operator as proposed in Refs.~\cite{Ghirardi1990,Tilloy2017}. {According to this prescription}, we substitute the mass density $\hat{\mu}(\mathbf{x})$ with the smeared one
\begin{equation}
\hat{\nu}(\mathbf{r})=\int \dd \mathbf{x} \,g(\mathbf{x}-\mathbf{r})\hat{\mu}(\mathbf{x}),
\end{equation}
where $g(\mathbf{x}-\mathbf{y})$ is a suitable smearing function. This is equivalent to {regularizing} both the noise kernel $\gamma(\mathbf{x}-\mathbf{y})$ and the  gravitational potential $\mathcal{V}(\mathbf{x}-\mathbf{y})$ with the same smearing function \cite{Tilloy2017}:
\begin{equation}
\gamma \rightarrow g \circ \gamma \circ g, \quad\text{and}\quad \mathcal{V} \rightarrow g \circ \mathcal{V} \circ g. \label{smearingcond}
\end{equation}

An appropriate smearing function should remove all the divergences of the master equation \eqref{TDmastereq} for an arbitrary choice of the mass density and of the noise kernel. 
In particular, $\hat{H}_{\text{grav}}$ in Eq.~\eqref{Hgrav} becomes 
\begin{equation}
\hat{H}'_{\text{grav}}=\frac{1}{2}\int \dd \mathbf{x} \dd \mathbf{y} \,\left(g \circ \mathcal{V} \circ g\right)(\mathbf{x}-\mathbf{y})\hat{\mu}(\mathbf{x})\hat{\mu}(\mathbf{y}),
\end{equation} 
and the decoherence kernel defined in Eq.~\eqref{decohker} turns into
\begin{equation}
D'(\mathbf{x}-\mathbf{y})=\left[\frac{g \circ \gamma \circ g }{8 \hbar^2}+\frac{1}{2}g \circ \left(\mathcal{V} \circ \gamma^{-1} \circ \mathcal{V} \right) \circ g \right](\mathbf{x}-\mathbf{y}).
\end{equation}
By substituting $\hat{H}_{\text{grav}}$ with $\hat{H}'_{\text{grav}}$ and $D(\mathbf{x}-\mathbf{y})$ with $D'(\mathbf{x}-\mathbf{y})$ in Eq.~\eqref{TDmastereq}, we obtain
\bqali
\frac{\dd \hat{\rho}}{\dd t}=&-\frac{i}{\hbar}\left[\hat{H}_0 + \hat{H}'_{\text{grav}},\hat{\rho} \right]\\
&-\int \dd \x \dd \y D'(\x-\y)\left[\hat{\mu}(\x),\left[\hat{\mu}(\y),\hat{\rho}\right]\right].
\label{TDtrue}
\eqali
In this way, we are able to retrieve a smeared quantum {Newtonian} gravitational interaction.

As a case of interest, we consider the  noise correlation function  by Eq.~\eqref{corrdel}. In such a case, we obtain
\bqali
\left(g \circ \mathcal{V} \circ g\right)(\mathbf{x}-\mathbf{y}) &= -4 \pi G \hbar^2 \eta_2(\mathbf{x}-\mathbf{y}),\\
\left(g \circ \gamma \circ g \right)(\mathbf{x}-\mathbf{y})&= A \eta_0(\mathbf{x}-\mathbf{y}),\\ 
\left[g \circ \left(\mathcal{V} \circ \gamma^{-1} \circ \mathcal{V} \right) \circ g \right](\mathbf{x}-\mathbf{y})&=\frac{16 \pi^2 G^2 \hbar^4}{A}\eta_4(\mathbf{x}-\mathbf{y}),
\label{coefficients}
\eqali
where we defined
\begin{equation}
\eta_n(\mathbf{x}-\mathbf{y})=\int \frac{\dd \mathbf{k}}{k^n} \,\tilde{g}^2(\mathbf{k}) e^{\frac{i}{\hbar}\mathbf{k} \cdot (\mathbf{x}-\mathbf{y})}, \label{eta}
\end{equation}
with $\tilde{g}(\mathbf{k})$ denoting the Fourier transform of $g(\mathbf{x}-\mathbf{y})$. A good smearing function {must give} finite expressions in Eq.~\eqref{coefficients}, which reflect an appropriate short-distance regularization of the {Newtonian} gravitational potential $\mathcal{V}(\mathbf{x}-\mathbf{y})$, the correlation kernel $\gamma(\mathbf{x}-\mathbf{y})$ and the feedback dynamics $\left(\mathcal{V} \circ \gamma^{-1} \circ \mathcal{V} \right)(\x-\y)$. In turn, one can exploit Eq.~\eqref{coefficients} to restrict the class of  smearing function. In particular, the requirement of the convergence of $\eta_4$ prevents the use of some intuitive choices for the smearing. Indeed, if one considers a Gaussian smearing 
$g(\mathbf{z})=\left(2 \pi \sigma^2\right)^{-3/2} \exp \left(-\z^2 /2 \sigma^2 \right)$, one has that both $\eta_0(\x-\y)$ and $\eta_2(\x-\y)$ converge, while $\eta_4(\mathbf{x}-\mathbf{y})$, in spherical coordinates,  becomes:
\begin{equation}
\eta_4(\mathbf{x}-\mathbf{y})=\frac{4 \pi}{(2 \pi \hbar)^3} \int_{0}^{\infty} {\dd k}\,\frac{e^{-\sigma^2k^2/\hbar^2}}{k^{2}} \frac{\sin \left(\frac{k}{\hbar} |\mathbf{x}-\mathbf{y}| \right)}{\frac{k}{\hbar}|\mathbf{x}-\mathbf{y}|}, 
\end{equation}
which diverges, since the integrand is not well defined for $k \to 0$. 

In the following, we determine the convergence requirements for the coefficients {$\eta_n(\x-\y)$}. 
For the sake of simplicity, we consider only spherical smearing functions, i.e. $\tilde{g}(\mathbf{k})=\tilde{g}(k)$. In such a case, Eq.~\eqref{eta} simplifies to
\begin{equation}
\eta_{n}(\mathbf{x}-\mathbf{y})=4 \pi \int_{0}^{\infty} \frac{\dd k}{k^{n-2}}\,\tilde{g}^2(k) \frac{\sin \left(\frac{k}{\hbar} |\mathbf{x}-\mathbf{y}| \right)}{\frac{k}{\hbar}|\mathbf{x}-\mathbf{y}|}, \label{generaleta}
\end{equation}
which {converges, for example,} for smearing functions of the family $\tilde{g}(k)=k^\beta e^{-\alpha k^2}$ with $\alpha > 0$ and $ \beta \geq 1$.
Concretely, a smearing function of the form
\begin{equation}
g(\mathbf{x}-\mathbf{y})=\frac{1}{(2 \alpha \hbar)^{7/2}}\left[ 6 \alpha \hbar^2 -(\mathbf{x}-\mathbf{y})^2 \right]e^{-\frac{(\mathbf{x}-\mathbf{y})^2}{4 \alpha \hbar^2}},
\end{equation}
whose Fourier transform {is}
\begin{equation}
\tilde{g}(k)=k^2e^{-\alpha k^2},
\end{equation}
belongs to such a family. In particular, explicit calculations lead to
\bqali
\eta_0 (\z)&= \frac{\pi^{3/2} }{16 \hbar^4 (2\alpha)^{11/2}} \left[\z^4+{40 \alpha \hbar^2}\left({6 \alpha \hbar^2} -\z^2\right) \right]e^{-\frac{\z^2}{8 \alpha \hbar^2} },\\
\eta_{2} (\z)&= -\frac{\pi^{3/2}}{4\hbar^2(2 \alpha)^{7/2}}\left(\z^2-{12 \alpha \hbar^2}\right) e^{-\frac{\z^2}{8 \alpha \hbar^2} },\\
\eta_{4}(\z)&=\frac{\pi^{3/2}}{(2 \alpha)^{3/2}}e^{-\frac{\z^2}{8 \alpha \hbar^2} },
\eqali
which are well defined also for $|\z|=|\mathbf{x}-\mathbf{y}|\to0$. In this way, the divergences in the TD model are indeed avoided.

If instead of  Eq.~\eqref{corrdel}, one takes $\gamma(\x-\y)=-2 \hbar \mathcal{V}(\x-\y)$, a normalized Gaussian smearing of standard deviation $\sigma$ leads to the following a decoherence kernel
\begin{equation}
D'(\x,\y)=\frac{G}{2 \hbar|\x-\y|}\operatorname{erf}\left(\frac{|\x-\y|}{2 \sigma} \right),
\end{equation}
which behaves well also for $|\x-\y|\to0$.\\

\end{document}